# Topological membrane devices for terahertz on-chip photonics

Quanlong Yang[1,*], Dongyang Wang[2], Sergey Kruk[1], Mingkai Liu[1], Ivan Kravchenko[3], Yuri Kivshar[1], and Ilya Shadrivov[1,*]

[1]*Nonlinear Physics Centre, School of Physics, Australian National University, Canberra ACT 2601, Australia*

[2]*Department of Physics, The Hong Kong University of Science and Technology, Clear Water Bay, Hong Kong 999077, China*

[3]*Center for Nanophase Materials Sciences, Oak Ridge National Laboratory, TN 37831, USA*

*Corresponding author. Email: quanlong.yang@anu.edu.au; ilya.shadrivov@anu.edu.au

**Abstract**

Terahertz waves offer a profound platform for next-generation sensing, imaging, and information communications. However, all conventional terahertz components and systems suffer from a bulky design, sensitivity to imperfections, and transmission losses. Here, we propose and experimentally demonstrate on-chip integration and miniaturization of topological devices which may address many existing drawbacks of the terahertz technology. We design and fabricate topological devices based on valley-Hall photonic structures that can be employed for various integrated components of on-chip terahertz systems. More specifically, we demonstrate the valley-locked asymmetric energy flow and mode conversion with topological straight waveguide, multi-port couplers, wave division, and whispering gallery mode resonators. Our devices are based on topological membrane metasurfaces which are of great importance for developing on-chip photonics and bring many novel features into terahertz devices.

**Introduction**

Terahertz photonics is well known as a powerful and efficient tool for biosensing, nondestructive imaging and detection of explosives[1, 2]. More recently, due to ever-increasing demands for broadband wireless communications, terahertz frequency range is considered to be a promising candidate for the next-generation communication technologies, including 6G systems and beyond[3]. On-chip devices whose dimensions are much smaller than free-space counterparts attract considerable interest. However, the existing terahertz devices are often bulky and lossy, and this limits their applicability in advanced technologies. The field of meta-optics offers pathways towards miniaturization and integration of photonic devices and systems[4, 5]. The precise control of both electric and magnetic resonances enabled by meta-optics is vital for developing efficient and compact electromagnetic components and devices[6-10]. Meta-devices based on free-standing dielectric membranes structure at the subwavelength scale are of particular practical importance as they remove many limitations associated with the negative impact of substrates on devices' performances[11-13].

Importantly, terahertz devices can further be empowered by the physics of topological phases. Indeed, the recent discoveries of photonic topological insulators (PTIs) reveal that such materials may support robust edge states being immune to disorder and sharp bends[14-16]. Numerous theoretical predictions and experimental demonstrations have been made in a broad spectral range, from microwaves to visible, as well as for two- and three-dimensional systems[17-23]. Early approaches to the topological phases rely on an external magnetic field or a careful design which hinders the widespread implementation. Meanwhile, great efforts have been put into the study of valley-Hall photonics crystals (VHPCs) or spin-Hall photonics crystals, which support robust edge modes inside a photonic bandgap, but they do not require time-reversal symmetry breaking [24-32].

For electromagnetic wave propagation, back-reflections is a major source of loss and feedback that hinders large-scale optical integration. While PTIs are strictly nonreciprocal and support one-way propagation, VHPCs are Pseudo spin-locked and support valley-polarized asymmetric propagation. The edge states of VHPCs propagate along the lattices interface, which provides a compact and reliable platform for on-chip integration[33, 34]. The topologically protected edge states of VHPCs are robust against defects and exhibit near-unity transmission even through sharp corners. Hence, on-chip VHPCs show promise for robust, low-loss and ultra-compact waveguides, which are crucial for future THz integrated photonics[34-36]. In addition, these waveguides are single-mode and exhibit nearly ideal linear dispersion properties. As a result, there is a growing interest in bringing VHPCs to on-chip integration, however, the practical topological devices for integrated on-chip terahertz manipulation still remain to be developed.

Here, we propose and realize experimentally several topology-empowered devices based on VHPCs of a perforated dielectric membrane for nontrivial on-chip terahertz wave manipulation. We demonstrate two types of domain walls separating rhombus lattices, where robust single-mode topological valley-polarized edge states are supported. In our experiment, in-plane electric field profiles of the topological waveguide are characterized with the terahertz near-field scanning system. In addition to the topological waveguide, we explore the topological multi-port coupler, wave splitter and whispering gallery mode ring resonator based on two types of domain walls. The designed dielectric membrane devices present additional degrees of freedom for photonic topological state manipulation in comparison with conventional dielectric waveguides. Our study paves the way for the development of a robust and efficient platform for terahertz on-chip communication, sensing, and multiplexing systems.

**Results and Discussion**

Figure 1a illustrates the schematic design of topological membrane devices: a low-index (air) circular hole pairs of radii of $R_a$ and $R_b$ are cut in a high-index (silicon) wafer. The unit cells are arranged in a rhombus lattice with spacing $a$ = 170 μm, and thickness of the suspended Si membrane is $h$ = 90 μm. The lattice vectors are shown as $a_1 = (a, 0)$ and $a_2 = (a/2, \sqrt{3}a/2)$. The radius difference of the two sublattice holes is denoted as $\triangle = |R_a - R_b|$. Here, we focus on the transverse-electric (TE)-like mode whose electric field lies predominantly in the $x$-$y$ plane.

We conduct the eigenmode analysis of the periodic structure using finite-element numerical method and extract the band diagram and field profiles as shown in Figure 1b (see Methods for simulation details). In the case with the inversion symmetry ($R_a = R_b = 0.2a$), a degeneracy of Dirac points (K valley) at 0.5 THz is shown by the dashed line in Figure 1b. By breaking the inversion symmetry ($R_a = 0.25a$, $R_b = 0.15a$), the Dirac point degeneracy is lifted, and a photonic band gap opens up from 0.46 THz to 0.54 THz (solid line). The in-plane electric field magnitude profile and electric field vectors of eigenmodes in the *K* valley are shown in the inset in Figure 1b. Two spin-polarized transverse modes are found for upper and lower bands, respectively, which correspond to the left-circular polarization (LCP) and right-circular polarization (RCP). Because they are not real spin polarizations, we call these LCP or RCP-like edge modes as pseudo-spin modes.

The realization of VHPCs relies on the breaking of the inversion symmetry, and it features a non-vanishing Berry curvature accumulated around different valleys. At *K* or *K'* point, each valley displays a valley-Chern invariant of $C_{k, k'}=\pm 1/2$ (pseudospin). A domain wall formed between two parity-inverted VHPCs possesses opposite valley-Chern numbers at a particular valley, which supports valley-polarized edge states within the optical bandgap. To study the edge states at the domain wall, we consider a waveguide configuration and numerically calculate the band diagram projection for a 20×1 supercell. The results are shown in Figure 1c, where two edge modes show up in the projection band gap, as indicated with red and blue curves, respectively. These two edge modes are found to be localized at the two types of domain walls DW1 (interface of (+Δ, - Δ) with larger holes adjacent to each other, edge mode in red curve) and DW2 (interface of (-Δ, + Δ) with smaller holes adjacent to each other, edge mode in blue curve) as shown in Figures 1d to 1e. In Figure1c, we denote the *K* valley edge mode at DW1 (or *K'* at DW2) as σ+, and *K'* valley edge mode at DW1 (or *K* at DW2) as σ-, indicating the direction of group velocity. We note that for a particular domain wall, the edge modes at each valley have opposite group velocities, which indicates the valley-locked edge modes and asymmetric propagation in the real space.  The dispersion of edge states is single-mode and almost linear within the two valleys, which provides uniform group velocity for achromatic propagation in the topological waveguide.

To confirm the edge mode dispersion properties, we performed the Fourier transformation of the in-plane electric field along the domain wall excited by a point dipole source. As displayed in Figures 1f and 1g, the in-plane electric field distributions are also calculated for two types of domain walls at 0.5 THz. It can be seen that for the forward propagating wave at DW1, only the $K$ valley is highlighted, which indicates that the excited edge state is $K$ valley locked (denoted as σ+). Meanwhile, for the forward propagating mode along the DW2, it is $K'$ valley locked (also denoted σ+), which shows an inverse valley-polarization compared to DW1.

To experimentally demonstrate the transmission signature and electric field profile of VHPCs, we fabricated the topological waveguide starting with a silicon-on-insulator wafer and using photolithography and deep reactive ion etching techniques (see Methods for fabrication details). Figures 2a and b display the optical and SEM image of the fabricated samples with DW1, and the white dashed line denotes the position of the interface that acts as the topological straight waveguide. This topological waveguide supports two counter-propagating edge modes with different valley-polarizations, for simplicity, only the forward propagating (σ+, along the positive $x$-direction) edge mode is considered here. Figure 2c shows the calculated transmission spectra of the topological waveguide (solid line), where an electric dipole placed on the surface of the membrane is used to excite the edge state in the simulation. A distinct transmission window from 0.46 THz to 0.54 THz can be clearly seen from the normalized transmission spectra for all the $E_x$, $E_y$ and $H_z$ field components.

In the experiment, a dual probe terahertz near-field scanning spectroscopy is employed to characterize the performance of topological waveguide (see Methods for experiment details), and the optical image of the terahertz emitter and detector is shown in Figure 2a. The employed system includes a near-field excitation source and a near-field detection probe (Protemics GmbH). Such configuration allows us to selectively excite individual waveguides and map field distribution around them. The measured transmission spectra are shown in Figure 2c as the dashed line for $E_y$ field component and it is in a good agreement with the calculated results. We attribute the discrepancy between the experiment and the calculated results to the fabrication imperfections, such as the deviation of the Si membrane thickness from design. The measured thickness of the fabricated samples is approximately 85 μm (the design thickness is 90 μm) and this thickness difference introduces the red shift of the transmission window. In addition, the discrepancy of the intensity comes from a strong dependency of the measured signal on the exact position of the detector. With a finite precision of the positioning of near-field antennas, the amplitude of the electric field varies significantly with the probe position above the sample due to the exponential decay of edge mode along the y-direction shown in Figure 2g.

In Figures 2d and 2e, we show the in-plane electric field ($E_x$, $E_y$) profile from simulation for 0.5 THz, where the TE-like field profile is presented. The edge modes are excited by the dipole source placed near the left edge of the sample and propagate along the x-direction, the field is confined to the straight waveguide, and it penetrates into each lattice by approximately five periods (850 μm). Figure 2f shows the experimentally measured $E_y$ component of electric field, a rectangular region marked by the black rectangle in Figure 2e is selected for the detailed experimental scan. The measured field in Figure 2f demonstrates a good agreement with simulation results. Due to the limitations of the THz near-field microscope, we are not able to measure the field distribution over the whole device. To show the good agreement of simulated and measured results, we present 1D electric field profiles across the waveguide and along the waveguide (lines 1 and 2 in Figure 2e) and plot them in Figures 2g and 2h. The measured field profile of $E_y$ component is in a good agreement with calculated results in Figure 2e, which demonstrated the interface confinement and exponential decay of the edge states. Here, the calculated decay length for simulation and experiment results are 810 and 780 μm, respectively. Further calculations are discussed in the Supplementary information.

To demonstrate the capability of terahertz on-chip devices based on VHPCs, we studied topological multi-port couplers. Due to the availability of the two types of domain walls, there are multiple combinations that can be used for constructing couplers, and we consider several of them. The surface modes at the domain walls are valley-locked as indicated in Figure 3a, and this prohibits the inter-valley scattering on a fixed domain wall and thus supports the possibility to build various directional couplers. Due to the inverted configuration of DW1 and DW2, the K(K') valley of DW1 is equivalent to K'(K) valley for DW2. For convenience, we denote the surface modes at K valley of DW1 and K' valley of DW2 as σ+, while surface mode at K' valley of DW1 and K valley of DW2 as σ-. The valley locking of surface states can thus be recognized as conservation of pseudospin σ, i.e., σ+(-) can only scatter into σ+(-).

We first consider a configuration of multi-port couplers shown in Figures 3b and 3c, where DW1 and DW2 form sharp bends and join at the center of the devices (DW1 and DW2 are marked by red and navy lines, respectively). For this topological four-port coupler, the supported propagation directions of σ+ surface mode are illustrated in Figure 3b. Port 1 with DW1 only supports the forward propagation of the σ+ edge state (K valley, blue arrow). Therefore, when the four-port coupler is excited from the port 1, the edge state σ+ propagates to the junction point. Due to the mismatch of the pseudospin, there is no transmission of energy to the port 3. Both ports 2 and 4 support coupling of the edge mode σ+, but the structure of the field distribution at the junction point and mode overlap between the waveguides determines the coupling efficiencies for these two ports. Only a small fraction of energy is allowed to couple into port 4, and port 2 receives most of the energy. These predictions agree well with the calculated electric field profiles ($E_y$) of the four-port coupler at 0.5 THz, as shown in Figure 3g. Because $E_x$ component of the four-port couplers show the similar conclusion, we did not show it here for simplification. To further characterize the propagation of the other edge states along with four ports, the calculated electric field profiles with the source placed in the ports 1 and 3 are presented in Figures 3h to 3j. When we excite other ports, the behavior of the coupler in these case shares the same features as the first case shown in the Figure 3g, which is in good agreement with the prediction.

We fabricate the four-port coupler as displayed in the inset of Figure 3g, where the SEM image of the junction point is shown. The measurement of four ports is taken 10 unit cells away from the junction point, which is marked by the points in Figure 3k with different colors, the measured transmission for four ports is shown in Figure 3k, where port 1 and port 2 display distinct transmission windows indicating the presence of edge modes within the bandgap (100% and 67% at 0.5 THz). Port 3 displays close to zero transmission (4.5%), and there is a fraction of energy that goes into port 4 (23%) as expected. There is only an approximately 5.5% transmission loss after the mode passes sharp corners, which demonstrates the robustness of our topological devices. Meanwhile, the mismatch of experimental and calculated transmission spectra comes from the fabrication error and low signal-to-noise ratio of the near-field setup. Moreover, the 1D electric field profiles show similarity to the measured results for the topological waveguide in Figure 2g, see the Supplementary information for further details.

We introduce another topological four-port coupler that is formed by only the DW1 type waveguide, as shown in Figures 3d and 3e. This coupler works as a traditional *directional coupler*. When the source is located at port 1, only the edge state σ+ is excited and the wave propagates towards ports 2 and 4, while nothing reaches port 3 due to the principle of pseudospin conservation. On the other hand, when excited from the port 4 (green color), the energy is coupled to ports 1 and 3. Excitation from ports 2 and 3 is shown in Figure 3e, where the reverse operation of the directional coupler can be seen. More excitation configurations and the demonstration of the valley-locked directional coupling can be found in the Supplementary information. Furthermore, the schematic of the topological six-port coupler consisting of six identical DW1 is shown in Figure 3f, which is equivalent to a combination of two directional couplers in Figures 3d and 3e. Similar to the earlier analysis of directional coupler, we find that there are four ports that support the valley-locked edge state with σ+ or σ-. For example, port 1 to ports 2, 4, 6 for σ+ and port2 to port 1, 3, 5 for σ-. The demonstration of the performance of the topological six-port couplers is presented in the Supplementary information.

Next, we demonstrate the topological beam-splitter which is shown in Figure 4. Figures 4a shows the schematics of the proposed topological beam-splitter. Three DW1 (red line) and three DW2 (navy line) are used in this device. When the splitter is excited from port 1, the forward propagating edge state σ+ is coupled into ports 2 and 3 with the ratio of 50:50. We show the calculated 2D electric field profile of the topological beam-splitter at 0.5 THz in Figure 4b, where the $E_y$ field components at ports 2 and 3 are evenly distributed. Zero transmission is found for port 4. We also fabricated the topological beam-splitter, the corresponding optical and SEM images of the fabricated sample are shown in Figures 4d and 4e. In Figures. 4c and 4f, we show the frequency dependency of the fabricated beam splitter with both simulation and experimental results. We note that the experimental curve appears to be smoother than the calculated one since the frequency resolution of our spectrometer is not sufficient for confirming all the fine features seen in the simulations. At points marked in Figure 4b with different colors, the transmission is studied, and we can see in Figure. 4c that the edge mode of port 1 (whose energy we take as 100% at 0.5 THz) is split into ports 2 and 3 with equal amplitude (48.3%), a close to zero transmission in port 4 (0.7%) and a 2.7% of transmission loss from the junction point is observed. The measured transmission for four ports is displayed in Figure 4f and it shows a good agreement with calculated results. Here, if we assume measured signal in port 1 at 0.5 THz as 100 %, then ports 2 and 3 receive 44.75% and 40%, respectively. Considering the loss from the junction and the bends, the signal splitting in the device is still close to 50 %. Moreover, by extending this beam splitter with its mirror image, we can construct a topological Mach-Zehnder interferometer, which can be employed for on-chip sensing and phase-sensitive measurements. Further discussion is presented in the Supplementary information.

The optical ring resonators offering an enhanced light-matter interaction are appealing for many applications ranging from bio-sensing to quantum optical devices, as well as integrated on-chip systems[37-39]. Topological photonics was used to create ring resonators[40-44], however we perform the first experiments in the terahertz frequency range, and utilize the dual near-field probe that allows us to excite the system at an arbitrary position. To explore potential on-chip devices based on the valley-locked edge states, we implement a topological membrane ring resonator based on the whispering gallery mode (WGM), as illustrated in Figure 5. For the conventional WGM resonator shown in Figure 5a, both waveguide and ring support the forward and backward propagating modes. When the straight waveguide is excited at the frequencies not matching the resonant frequency of the WGMs, there will be near-zero coupling between the waveguide and the ring. When the frequencies match, the energy flows into the ring resonator and affects the transmission of the wave through the waveguide. The forward and backward coupling between the ring and the waveguide are both supported.

By alternately arranging DW1 and DW2, we built a topological hexagonal WGM resonator, which supports clockwise propagation of σ+ and counter-clockwise propagation of σ-. The propagation of edge modes at the hexagonal ring complies with the same mechanism of multi-port couplers in Figure 3a, and there is very small reflection at the bends. In Figure 5b, we show coupling of the waveguide mode to the topological hexagonal WGM resonator, when the waveguide away from the resonator is excited. Away from the resonant frequencies, the behavior of the topological resonator is similar to the conventional WGM resonator except for the suppression of the backward propagating mode (× of backward modes). For the resonator shown in Figure 5b, there is only counter-clockwise propagating mode σ+ that is allowed to couple to the waveguide. As to the coupling output, modes in the ring resonators due to pseudospin matching only support the coupling to the forward mode (√). More interestingly, when the critical coupling is satisfied, no edge state on the ring could couple forward, and the energy is trapped within the ring (× of forward modes).

Simulation results for the resonances of the topological WGM resonator are shown in Figures 5d and 5e, where the spacing between the resonator and the waveguide is equal to five periods (See the supplementary information). In Figure 5d, we excite from the waveguide and calculate the electric field at two points shown in the figure: blue is on the ring, and red is on the waveguide. There are three main resonance peaks 2-4 (0.518, 0.523, and 0.528 THz, the resonance frequency positions depend on the circumference of the ring) on the rings and three transmission dips in the waveguide field spectrum, which indicate that the energy is coupled to the ring. In particular, the transmission for 0.528 THz drops to near zero, which corresponds to the critical coupling regime. We show the mode structure for these discrete frequencies in Figure 5e. The simulation results show that no electric field couples to the ring at 0.5 THz, as the wave propagates along the straight waveguide. Due to the strong coupling at 0.518 and 0.523 THz, the power flows into the ring resonator and obvious traveling wave patterns are observed in the ring. As for coupling output, only the forward modes are supported, and they are marked by "√" in schematics in Figure 5b, and their direction is shown by the white arrow in Figure 5e. Moreover, the edge modes at 0.528 THz display a zero transmission, we can see that all the energy is trapped in the ring and the electric field profile shows a standing wave pattern. Due to the limited frequency resolution our current near-field time-domain spectroscope, we were not able to detect these high-Q resonances in the experiment.

We also study a different configuration, when we excite the topological WGM resonator directly, this is shown in Figure 5c. Since we use the point dipole source, the edge modes with different propagation directions (σ+, counter-clockwise, and σ- for clockwise) are both excited. The out-coupling from the ring to the waveguide is found in two situations. It is worth mentioning that the edge states σ+ and σ- couple to the waveguide in opposite directions. Figure 5f illustrates the fields in the resonator and waveguide when the resonator is excited directly. In comparison to the case of the waveguide excitation, due to the coupling limitation between waveguide and ring, the excitation of the ring directly shows larger number of WGMs. The electric field profiles of the edge states at four discrete frequencies are shown in the right panel of Figure 5g, where opposite out-coupling to the waveguide can be seen.

**Conclusion**

In summary, we have designed and demonstrated several terahertz topological devices based on VHPCs for integrated on-chip systems. Valley locked edge modes at discrete domain walls are utilized for mode coupling and conversion. The employment of a terahertz near-field spectroscopy system allows us to directly characterize and visualize the valley-locked edge states on our proposed topological devices, including waveguide, directional couplers, and beam splitters. We also studied the coupling of topological waveguides to topological ring resonators and observed selective coupling defined by the mode pseudo-spin. This demonstration opens an avenue towards terahertz topological on-chip photonic networks for next-generation wireless communications, bio-sensing, and quantum computing.

**Experimental section**

*Numerical simulation:* The eigenmode analysis is performed using COMSOL Multiphysics, where periodic boundaries are used in the simulations. In addition, CST Microwave Studio (the frequency-domain solver with open boundary conditions) is employed to characterize the transmission spectra and field distributions of topological membrane devices. Open boundary conditions were applied in all directions. The simulation results were obtained 10 μm above the upper surface of the topological devices in order to emulate experimental conditions, where the fields are mesured some distance above the surface. The discrete port is served as a point dipole source and its polarization is located in the *x-y* plane. All the topological devices are made of a free-standing silicon membrane with a permittivity of $\varepsilon = 11.7$. Field distributions of the edge modes were mapped by defining the electric field monitors at 0.5 THz. In the simulations, the transmission spectrum of the multi-port coupler and splitter are normalized to the maximum field in the port 1. In the case of the multi-port devices, the field amplitude in the waveguide leading to the port 1 at 0.5 THz is used to normalize the fields in the remaining ports.

*Fabrication:* We fabricate the metasurface on a silicon-on-insulator wafer with a high-resistivity 90-μm-thick silicon layer, 1-μm-thick buried oxide layer and a 300-μm-thick handle wafer. Both polished sides of the wafer are initially coated with 1-μm-thick silicon oxide. Standard photolithography is employed to delineate the pattern of the membrane metasurface in the photoresist film. After a hard bake, this photoresist layer is used to conduct a plasma etching process of the silicon oxide film which, in turn, served as the hard mask during deep reactive etching of the device layer. After the resist removal and surface cleaning, the resulting metasurface in the device layer is passivated by plasma-enhanced chemical vapor deposition of a 1.4 μm-thick silicon oxide to provide the device membrane with protection and mechanical support. Similar production steps are repeated to etch the windows in the handle layer after a back-side alignment procedure. As a result, there is a rectangular window in the handle wafer at the back of topological devices, which look as membranes suspended on the remaining ribs of the handle wafer. *Angle-resolved THz system*: Two fiber-coupled terahertz antennas are used as a transmitter and detector (See Figure S5 in the Supporting information). The antennas are mounted on two guide rails that can rotate concentrically with the rotator. The rotator controls the detection angle of the THz beam. The THz wave emitted from the transmitter is collimated by the lens to deliver a nearly collimated beam with a diameter of 5 mm. The samples are placed at the center of the rotator. The deflected terahertz beam is collected by a lens and received by the detector. The detection angle relative to the incident normal varied from -30° to 30° with a step of 2° for the membrane deflector measurements.

*Characterization*: In the experiment, a broadband fiber-based terahertz near-field spectroscopy system is employed to characterize the performance of the topological membrane devices. Two terahertz probes are used as the transmitter and detector, respectively. The emitter that serves as a dipole source is mounted on a three-dimension motorized translation stage which enabled 2D scans at a fixed distance from the sample surface. nly the in-plane polarization couples substantially to the topological modes. For the 2D scanning and transmission measurement, the detector probe is placed at either ~50 μm (for scans) or ~30 μm (for single point measurement) above the sample, respectively. Closer placement of the probe to the sample for single point measurements provided greater signal-to-noise ratio, while we had to resort to larger spacing for the scans in order to avoid crashing the probe into the sample due to non-ideally parallel planes of the sample and the scanning area. The 1D and 2D electric field were detected with 20 μm per step in both x and y directions.

**Acknowledgments**

The work was supported by the Australian Research Council. A portion of this research was conducted at the Center for Nanophase Materials Sciences, which is a DOE Office of Science User Facility.

**References**

1. Singh R, Cao W, Al-Naib I, Cong L, Withayachumnankul W, Zhang W. Ultrasensitive terahertz sensing with high-q fano resonances in metasurfaces. *Appl. Phys. Lett.* **105**, 171101 (2014).

2. Mittleman DM. Twenty years of terahertz imaging. *Opt. Express* **26**, 9417-9431 (2018).

3. Sarieddeen H, Saeed N, Al-Naffouri TY, Alouini M-S. Next generation terahertz communications: A rendezvous of sensing, imaging, and localization. *IEEE Commun. Mag.* **58**, 69-75 (2020).

4. Zheludev NI, Kivshar YS. From metamaterials to metadevices. *Nat. Mater.* **11**, 917-924 (2012).

5. Glybovski SB, Tretyakov SA, Belov PA, Kivshar YS, Simovski CR. Metasurfaces: From microwaves to visible. *Phys. Rep.* **634**, 1-72 (2016).

6. Yu N, *et al.* Light propagation with phase discontinuities: Generalized laws of reflection and refraction. *Science* **334**, 333 (2011).

7. Yang Y, Kravchenko II, Briggs DP, Valentine J. All-dielectric metasurface analogue of electromagnetically induced transparency. *Nat. Comm.* **5**, 5753 (2014).

8. Wang S, *et al.* Broadband achromatic optical metasurface devices. *Nat. Comm.* **8**, 187 (2017).

9. Kruk S, Kivshar Y. Functional meta-optics and nanophotonics governed by mie resonances. *ACS Photon.* **4**, 2638-2649 (2017).

10. Sun S, He Q, Hao J, Xiao S, Zhou L. Electromagnetic metasurfaces: Physics and applications. *Adv. Opt. Photon.* **11**, 380-479 (2019).

11. Ong JR, Chu HS, Chen VH, Zhu AY, Genevet P. Freestanding dielectric nanohole array metasurface for mid-infrared wavelength applications. *Opt. Lett.* **42**, 2639-2642 (2017).

12. Yang Q, *et al.* Mie-resonant membrane huygens' metasurfaces. *Adv. Func. Mater.* **30**, 1906851 (2020).

13. Yang Q, *et al.* Polarization-sensitive dielectric membrane metasurfaces. *Adv. Opt. Mater.* **8**, 2000555 (2020).

14. Khanikaev AB, Hossein Mousavi S, Tse W-K, Kargarian M, MacDonald AH, Shvets G. Photonic topological insulators. *Nat. Mater.* **12**, 233-239 (2013).

15. Lu L, Joannopoulos JD, Soljačić M. Topological photonics. *Nat. Photon.* **8**, 821-829 (2014).

16. Khanikaev AB, Shvets G. Two-dimensional topological photonics. *Nat. Photon.* **11**, 763-773 (2017).

17. Rechtsman MC, *et al.* Photonic floquet topological insulators. *Nature* **496**, 196-200 (2013).

18. Hafezi M, Mittal S, Fan J, Migdall A, Taylor JM. Imaging topological edge states in silicon photonics. *Nat. Photon.* **7**, 1001-1005 (2013).

19. Blanco-Redondo A, Bell B, Oren D, Eggleton BJ, Segev M. Topological protection of biphoton states. *Science* **362**, 568 (2018).

20. Barik S, *et al.* A topological quantum optics interface. *Science* **359**, 666 (2018).

21. Yang Y, *et al.* Realization of a three-dimensional photonic topological insulator. *Nature* **565**, 622-626 (2019).

22. Smirnova D, Kruk S, Leykam D, Melik-Gaykazyan E, Choi D-Y, Kivshar Y. Third-harmonic generation in photonic topological metasurfaces. *Phys. Rev. Lett.* **123**, 103901 (2019).

23. Wang D, *et al.* Photonic weyl points due to broken time-reversal symmetry in magnetized semiconductor. *Nat. Phys.* **15**, 1150-1155 (2019).

24. Dong J-W, Chen X-D, Zhu H, Wang Y, Zhang X. Valley photonic crystals for control of spin and topology. *Nat. Mater.* **16**, 298-302 (2017).

25. Noh J, Huang S, Chen KP, Rechtsman MC. Observation of photonic topological valley hall edge states. *Phys. Rev. Lett.* **120**, 063902 (2018).

26. Gao F, *et al.* Topologically protected refraction of robust kink states in valley photonic crystals. *Nat. Phys.* **14**, 140-144 (2018).

27. Shalaev MI, Walasik W, Tsukernik A, Xu Y, Litchinitser NM. Robust topologically protected transport in photonic crystals at telecommunication wavelengths. *Nat. Nanotech.* **14**, 31-34 (2019).

28. Wang M, Zhou W, Bi L, Qiu C, Ke M, Liu Z. Valley-locked waveguide transport in acoustic heterostructures. *Nat Commun* **11**, 3000 (2020).

29. Arora S, Bauer T, Barczyk R, Verhagen E, Kuipers L. Direct quantification of topological protection in symmetry-protected photonic edge states at telecom wavelengths. *Light Sci. Appl.* **10**, 9 (2021).

30. Muhammad Talal Ali Khan HL, Nathan Nam Minh Duong, Andrea Blanco-Redondo, Shaghik Atakaramians. 3d-printed terahertz topological waveguides. *arXiv preprint* **arXiv:2010.16299.**, (2020).

31. Ma T, Shvets G. All-si valley-hall photonic topological insulator. *New J. Phys.* **18**, 025012 (2016).

32. Xue H, Yang Y, Zhang B. Topological valley photonics: Physics and device applications. *Adv. Photon. Res.* **n/a**, 2100013 (2021).

33. Cha J, Kim KW, Daraio C. Experimental realization of on-chip topological nanoelectromechanical metamaterials. *Nature* **564**, 229-233 (2018).

34. Yang Y, *et al.* Terahertz topological photonics for on-chip communication. *Nat. Photon.* **14**, 446-451 (2020).

35. Xiong H, *et al.* Topological valley transport of terahertz phonon–polaritons in a linbo3 chip. *ACS Photon.* **8**, 2737-2745 (2021).

36. Zeng Y, *et al.* Electrically pumped topological laser with valley edge modes. *Nature* **578**, 246-250 (2020).

37. Yang Y, Hang ZH. Topological whispering gallery modes in two-dimensional photonic crystal cavities. *Opt. Express* **26**, 21235-21241 (2018).

38. Wang C, *et al.* Electromagnetically induced transparency at a chiral exceptional point. *Nat. Phys.* **16**, 334-340 (2020).

39. Yu S-Y, *et al.* Critical couplings in topological-insulator waveguide-resonator systems observed in elastic waves. *Natl. Sci. Rev.* **8**, (2021).

40. Bandres MA, *et al.* Topological insulator laser: Experiments. *Science* **359**, eaar4005 (2018).

41. Jalali Mehrabad M, *et al.* Chiral topological photonics with an embedded quantum emitter. *Optica* **7**, 1690 (2020).

42. Liang G, Chong Y. Optical resonator analog of a two-dimensional topological insulator. *Phys. Rev. Lett.* **110**, 203904 (2013).

43. Lin Q, Sun X-Q, Xiao M, Zhang S-C, Fan S. A three-dimensional photonic topological insulator using a two-dimensional ring resonator lattice with a synthetic frequency dimension. *Sci. Adv.* **4**, eaat2774 (2018).

44. Gu L, *et al.* A topological photonic ring-resonator for on-chip channel filters. *J. Light. Techno.* **39**, 5069-5073 (2021).

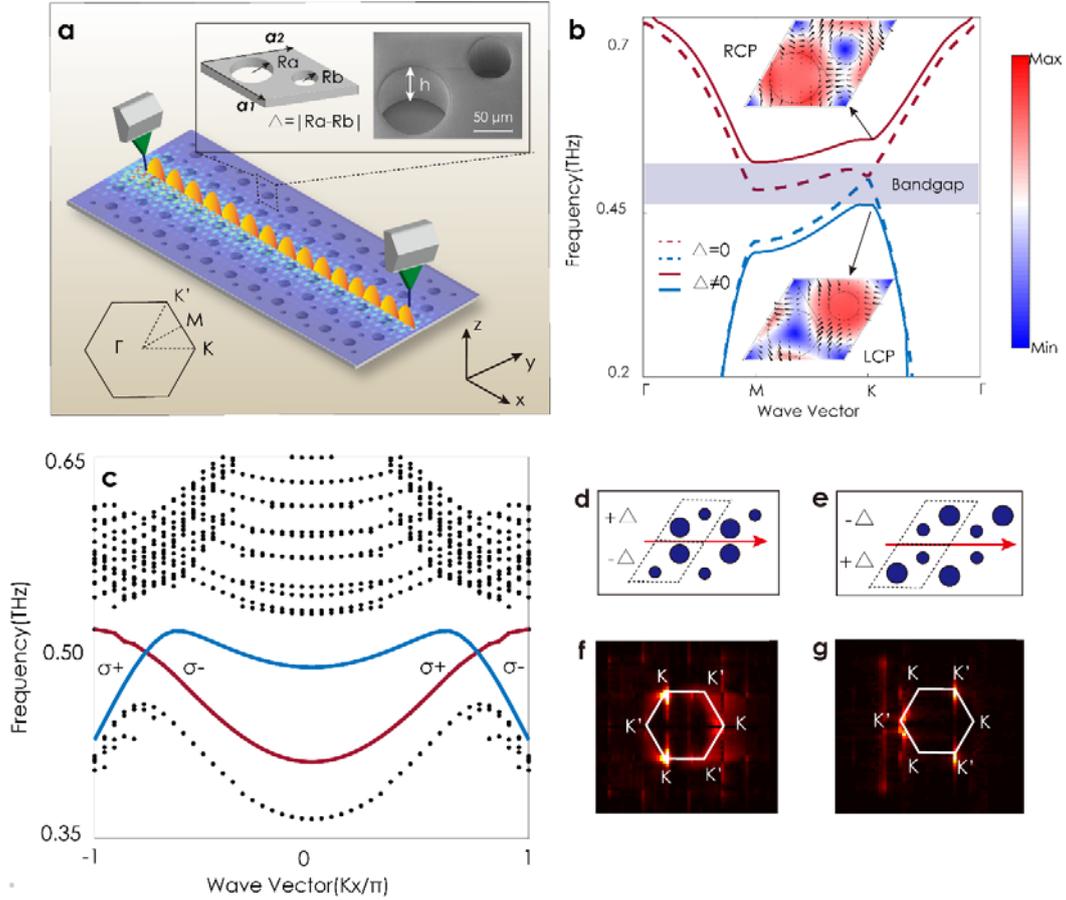

**Figure 1.** Concept of topological membrane meta-devices. (a) Schematic of topological membrane meta-devices and terahertz near-field scanning microscope systems. Two terahertz probes are employed as the emitter and detector to map the in-plane field distributions of topological meta-devices. Inset shows the schematic view and electron microscope image of the fabricated unit cell of the VHPCs. (b) Photonic band structure of meta-atoms with (dashed line) and without (solid line) inversion symmetry, $\triangle = |R_a - R_b|$ is the radius difference of the two circular holes. Inset shows the mode profiles of meta-atoms in K valley, and the arrows denote the electric field vectors. (c) Calculated dispersion of topological edge states from a 20×1 supercell, data points in black are projection bands. (d, e) Schematic of the two studied domain walls (DW1 and DW2). The $\triangle$ is of opposite sign for the two sides of domain walls, as is also seen as mirror arrangement of holes across the interface. DW1 and DW2 are denoted as (+$\triangle$, -$\triangle$) and (-$\triangle$, +$\triangle$), respectively. (f, g) Fourier spectra of DW1 and DW2 at 0.5 THz obtained from calculated real space electric field distribution.

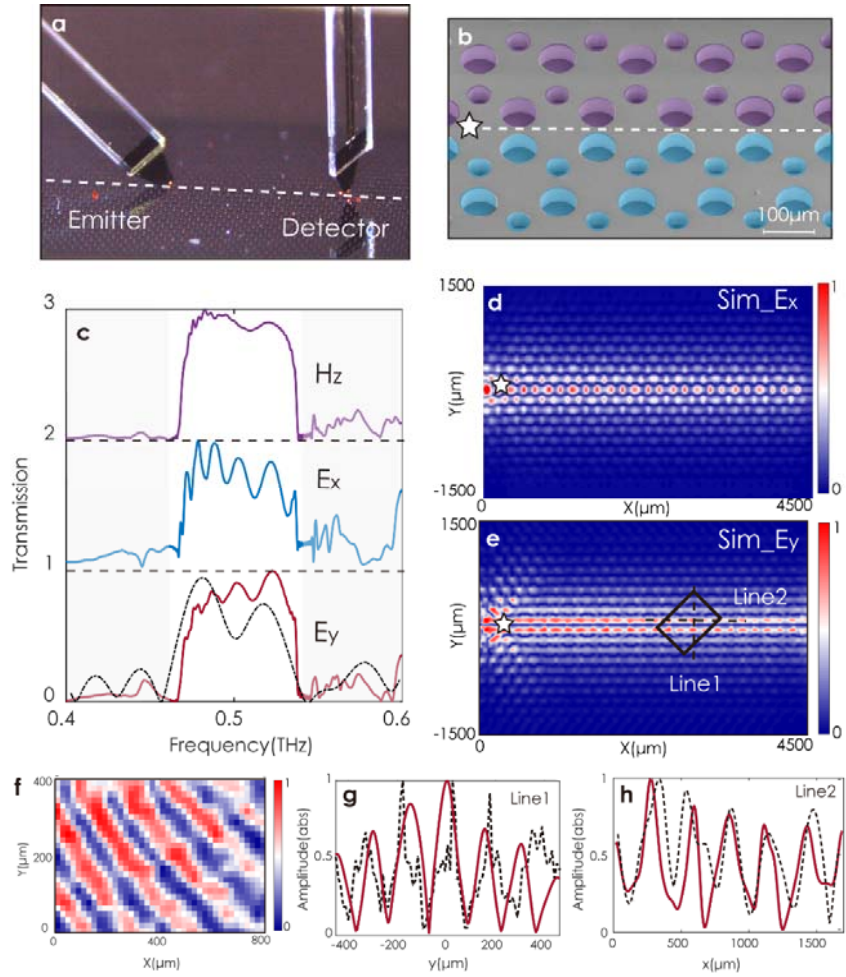

**Figure 2.** Topological on-chip waveguides. (a) Optical image of terahertz near-field microscope probes and fabricated topological waveguide. The white dashed line denotes the position of the topological waveguide as DW1. (b) SEM image of the fabricated straight waveguide with two parity-inverted VHPCs lattices. (c) Calculated (solid line) and measured (dashed line) transmission curves of $E_x$, $E_y$ and $H_z$ for the topological straight waveguide. (d, e) Calculated in-plane electric field magnitude distribution for the straight topological waveguide at 0.5 THz. The white star denotes the position of the excitation source. (f) Measured Ey field distributions of topological waveguide marked by the black rectangle in (e). (g, h) Calculated (red solid line) and measured (black dashed line) 1D $E_y$ field distributions across and along the waveguide (lines 1 and 2 in (e)), respectively.

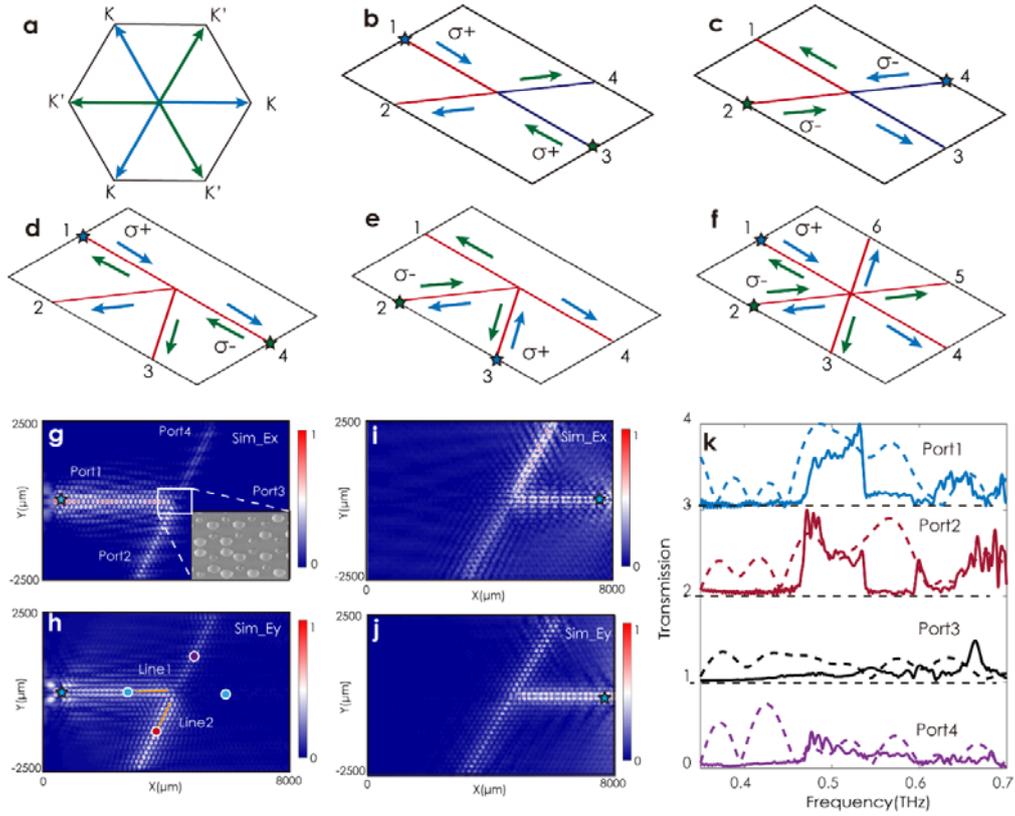

**Figure 3.** Topological on-chip multi-port couplers. (a) Valley-locked surface modes at domain walls. The blue and green arrows denote surface mode at K and K' valley, respectively. (b, c) Schematic of the four-port directional coupler consisting of DW1 (red) and DW2 (navy). The arrows denote the power flow of two edge states at K and K' valley, and the stars indicate the excitation source. (d-f) Conceptual illustration of the topological directional coupler with four identical DW1 and six-port coupler consists of six identical DW1. (g, h) Calculated in-plane field distribution ($E_x$ and $E_y$) on the topological four-port coupler at 0.5 THz with the source at port 1. Inset is the SEM image of the junction point. (i, j) Correponding calculated in-plane field distribution with the source at port 3. (k) Calculated and measured transmission of four different ports with the excitation from port 1. Each measurement is taken 10 unit cells from the junction so that the measurement at port 1 corresponds to the incident wave. The signal detection positions are shown by colored circles in (h).

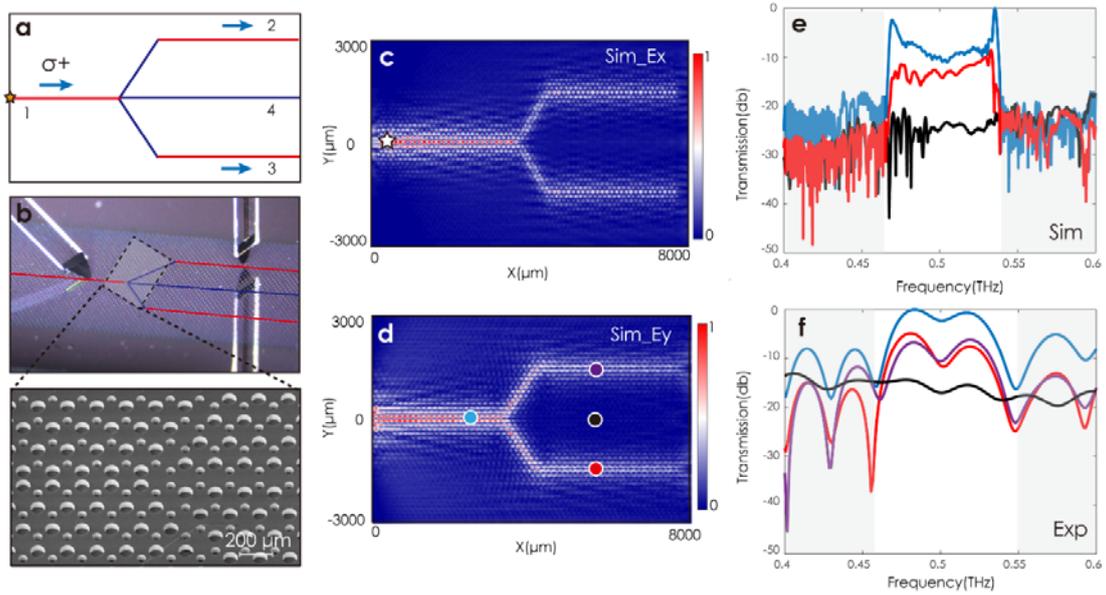

**Figure 4.** Topological on-chip splitters. (a) Schematics of the topological splitter containing DW1 (red line) and DW2 (navy line) waveguides. The blue and green arrows denote the power flow of the two edge states σ+ and σ-. (b) Optical image of the fabricated topological beam-splitter placed in the terahertz near-field microscope. Inset shows the SEM image of the junction point. (c, d) Calculated in-plane electric field distribution on the topological splitter at 0.5 THz. (e, f) Calculated and experimentally measured transmission spectra for in-plane components at positions 1-4. The signal detection positions are shown in (d).

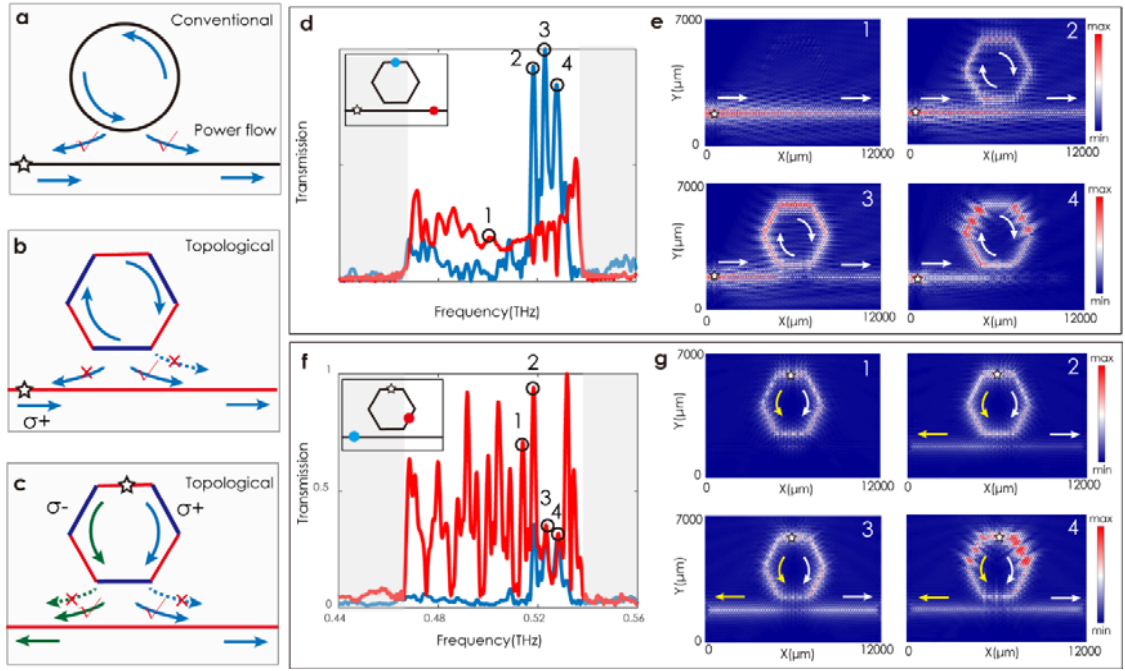

**Figure 5.** Topological on-chip resonators. (a) Mode coupling of conventional whispering gallery mode resonator. The arrows show the direction of possible power flow, while stars indicate the position of the excitation source. The energy from the ring couple into both forward and backward propagating modes of the waveguide. (b) Mode conversion of topological resonators with the source located in the waveguide. The arrows denote the supported (√) and forbidden (×) directions of coupling, and red and navy lines denote DW1 and DW2, respectively. Two types of resonances (traveling and standing waves) arise depending on the coupling of the forward propagation. The dashed arrow denotes the critical coupling. (c) Corresponding mode conversion of topological resonators with the source located on the ring. (d) Electric field spectrum in the resonator (measured at the location of the blue point shown in the inset) and straight waveguide (red point). (e) $E_y$ field distributions of topological resonators at peaks 1-4 (0.5, 0.518, 0.523, and 0.528 THz). The white arrows denote the power flow of the edge modes σ+. (f, g) Corresponding electric field spectrum and field distributions of four peaks 1-4 (0.513, 0.518, 0.523, and 0.528 THz) with the source exciting the ring directly. The white and yellow arrows display the power flow of the edge modes σ+ and σ-, respectively.